\documentclass[12pt,a4paper]{amsart}
\usepackage[T1]{fontenc}
\usepackage{amssymb}
\usepackage{graphicx}
\usepackage[colorlinks=true,urlcolor=blue,citecolor=blue,linkcolor=blue,pdfstartview=FitH,pdfpagemode=None]{hyperref}
\hypersetup{
    pdftitle={Bi-Hamiltonian systems on the dual of the Lie algebra of vector fields of the circle and periodic shallow water equations}, 
    pdfauthor={Boris Kolev}, 
    pdfsubject={2000 Mathematics Subject Classification: 35Q35, 35Q53, 37K10, 37K65}, 
    pdfkeywords={Bi-Hamiltonian formalism, Diffeomorphisms group of the circle, Lenard scheme, Camassa-Holm equation} 
    }
\newtheorem{theorem}{Theorem}[section]

\newtheorem{lemma}[theorem]{Lemma}
\newtheorem{proposition}[theorem]{Proposition}

\theoremstyle{definition}
\newtheorem{definition}[theorem]{Definition}

\theoremstyle{remark}
\newtheorem{remark}[theorem]{Remark}
\newtheorem{example}[theorem]{Example}

\newcommand{\rd}{\mathrm{d}}
\newcommand{\ri}{\mathrm{i}}
\newcommand{\re}{\mathrm{e}}
\newcommand{\const}{\mathrm{const.}}

\newcommand{\g}{\mathfrak{g}}
\newcommand{\gstar}{\mathfrak{g}^{*}}
\newcommand{\VectS}{\mathrm{Vect}(S^{1})}
\newcommand{\VectSstar}{\mathrm{Vect}^{*}(S^{1})}
\newcommand{\Ucircle}{S^{1}}
\newcommand{\DiffS}{\mathrm{Diff}(S^{1})}
\newcommand{\CS}{C^{\infty}(S^{1})}
\newcommand{\CM}{C^{\infty}(M)}

\newcommand{\set}[1]{\left\{#1\right\}}

\newcommand{\brackets}[2]{\left\langle #1,#2\,\right\rangle}
\newcommand{\poisson}[2]{\{ #1,#2\,\}}
\newcommand{\dd}[2]{\frac{d{#1}}{d{#2}}}

\DeclareMathOperator{\tr}{tr} %
\begin{document}

\title[Bi-Hamiltonian systems]{Bi-Hamiltonian systems on the dual of the Lie algebra of vector fields of the circle and periodic shallow water equations}%
\author[B. Kolev]{Boris Kolev}%
\address{CMI, 39 rue F. Joliot-Curie, 13453 Marseille cedex 13, France}
\email{boris.kolev@cmi.univ-mrs.fr}

\thanks{This paper was written
during the author's visit to the Mittag-Leffler Institute in
October, 2005, in conjunction with the Program on Wave Motion. The
author wishes to extend his thanks to the Institute for its generous
sponsorship of the program, as well as to the organizers
for their work. The author expresses also his gratitude to David Sattinger for several remarks that helped to improve this paper.}%

\subjclass[2000]{35Q35, 35Q53, 37K10, 37K65}%
\keywords{Bi-Hamiltonian formalism, Diffeomorphisms group of the circle, Lenard scheme, Camassa-Holm equation}%

\date{16 ao\^{u}t 2006}%
\begin{abstract}
This paper is a survey article on bi-Hamiltonian systems on the dual
of the Lie algebra of vector fields on the circle. We investigate
the special case where one of the structures is the canonical
Lie-Poisson structure and the second one is constant. These
structures called \emph{affine} or \emph{modified Lie-Poisson
structures} are involved in the integrability of certain Euler
equations that arise as models of shallow water waves.
\end{abstract}
\maketitle

\section{Introduction}
\label{sec:intro}

In the last forty years or so, the Korteweg-de Vries equation has
received much attention in the mathematical physics literature. Some
significant contributions were made in particular by Gardner, Green,
Kruskal, Miura (see \cite{PS05} for a complete bibliography and a
historical review). It is through these studies, that emerged the
\emph{theory of solitons} as well as the \emph{inverse scattering
method}.

One remarkable property of Korteweg-de Vries equation, highlighted
at this occasion, is the existence of an infinite number of first
integrals. The mechanism, by which these conserved quantities were
generated, is at the origin of an algorithm called the \emph{Lenard
recursion scheme} or \emph{bi-Hamiltonian formalism}
\cite{GD79,Mag78}. It is representative of infinite-dimensional
systems known as \emph{formally integrable}, in reminiscence of
finite-dimensional, classical integrable systems (in the sense of
Liouville). Other examples of bi-Hamiltonian systems are the
Camassa-Holm equation \cite{FF81,CH93,Con98,CM99,GH03} and the Burgers
equation.

One common feature of all these systems is that they can be
described as the geodesic flow of some right-invariant metric on the
diffeomorphism group of the circle or on a central real extension of
it, the Virasoro group. Each left (or right) invariant metric on a
Lie group induces, by a reduction process, a canonical flow on the
\emph{dual of its Lie algebra}. The corresponding evolution
equation, known as the \emph{Euler equation}, is Hamiltonian
relatively to some canonical \emph{Poisson structure}. It
generalizes the Euler equation of the free motion of a rigid
body\footnote{In that case, the group is just the rotation group,
$SO(3)$.}. In a famous article~\cite{Arn66}, Arnold pointed out that
this formalism could be applied to the group of volume-preserving
diffeomorphisms to describe the motion of an ideal fluid\footnote{
However, this formalism seems to have been extended to hydrodynamics
before Arnold by Moreau \cite{Mor59}.}. Thereafter, it became clear
that many equations from mathematical physics could be interpreted
the same way.

In \cite{GD81} (see also \cite{OK87}), Dorfman and Gelfand showed
that Korteweg-de Vries \cite{KdV95} equation can be obtained as the
geodesic equation, on the Virasoro group, of the right-invariant
metric defined on the Lie algebra by the $L^2$ inner product. In
\cite{Mis98}, Misiolek has shown that Camassa-Holm equation
\cite{CH93} which is also a one dimensional model for shallow water
waves, can be obtained as the geodesic flow on the
Virasoro group for the $H^{1}$-metric.

While both the Korteweg-de Vries and the Camassa-Holm equation have a geometric derivation and both are models for the propagation of shallow water waves, the two equations have quite different structural properties. For example, while all smooth periodic initial data for the Korteweg-de Vries equation develop into periodic waves that exist for all times \cite{Tao02}, smooth periodic initial data for the Camassa-Holm equation
develop either into global solutions or into breaking waves (see the papers \cite{Con97,CE98b,CE00,McK04}).

In this paper, we study the case of right-invariant metrics on the
diffeomorphism group of the circle, $\DiffS$. Notice however that a similar theory is likely
without the periodicity condition (in which case, some weighted spaces express how close the diffeomorphisms of the line are to the identity \cite{Con00}).

Each right-invariant metric on $\DiffS$ is
defined by an inner product $\mathbf{a}$ on the Lie algebra of the
group, $\VectS = \CS$. If this inner product is \emph{local}, it is
given by the expression
\begin{equation*}
    \mathbf{a}(u,v) = \int_{\Ucircle} u\, A(v) \, dx \qquad u,v \in \CS,
\end{equation*}
where $A$ is an invertible, symmetric, linear differential operator.
To this inner product on $\VectS$, corresponds a quadratic
functional (the energy functional)
\begin{equation*}
    H_{A}(m) = \frac{1}{2}\int_{\Ucircle} m\, A^{-1}(m),
\end{equation*}
on the (regular) dual $\VectSstar$. Its corresponding Hamiltonian
vector field $X_{A}$ generates the Euler equation
\begin{equation*}
    \dd{m}{t} = X_{A}(m).
\end{equation*}

Among Euler equations of that kind, we have the well-known
\emph{inviscid Burgers} equation
\begin{equation*}
    u_{t} + 3uu_{x} = 0,
\end{equation*}
and \emph{Camassa-Holm} \cite{CH93,FF81} shallow water equation
\begin{equation*}
    u_{t} + uu_{x} + \partial_{x} \, (1 - \partial_{x}^{2})^{-1}\Bigl( u^{2} + \frac{1}{2} \, u_{x}^{2} \Bigr) = 0.
\end{equation*}
Indeed, the inviscid Burgers equation corresponds to $A = I$
($L^{2}$ inner product), whereas the Camassa-Holm equation
corresponds to $A = I - D^{2}$ ($H^1$ inner product) (see
\cite{CK02,CK03}).

Burgers, Korteweg-de Vries and of Camassa-Holm equations are
precisely bi-Hamiltonian relatively to some second \emph{affine}
(after Souriau \cite{Sou97}) compatible Poisson
structure\footnote{The affine structure on the Virasoro algebra
which makes Korteweg-de Vries equation a bi-Hamiltonian system seems
to have been first discovered by Gardner \cite{Gar71} and for this
reason, some authors call it the \emph{Gardner bracket} (see also
\cite{FZ71}.} (see \cite{CM99,Len04b,McK79}). Since these equations
are special cases of Euler equations induced by $H^{k}$-metric, it
is natural to ask whether, in general, these equations have similar
properties for any value of $k$. In \cite{CK06}, it was shown that
this \emph{was not the case}. There are no affine structure on
$\VectSstar$ which makes the Eulerian vector field $X_{k}$,
generated by the $H^{k}$-metric, a bi-Hamiltonian system, unless
$k=0$ (Burgers) or $k=1$ (Camassa-Holm). One similar result for the
Virasoro algebra was given in \cite{CKL06}. We investigate, here,
the problem of finding a modified Lie-Poisson structure for which
the vector field $X_{A}$ is bi-Hamiltonian. We show, in particular,
that for an operator $A$ with constant coefficients, this is
possible only if $A = aI + bD^{2}$, where $a,b \in \mathbb{R}$.

In \S\ref{sec:Poisson}, we recall the definition of Hamiltonian and
bi-Hamiltonian manifolds and the basic materials on bi-Hamiltonian
vector fields. Section~\ref{sec:Lie_Poisson} contains a description
of Poisson structures on the dual of the Lie algebra of a Lie group.
The last section is devoted to the study of bi-Hamiltonian Euler
equations on $\VectSstar$; the main results are stated and proved.

In the description of modified affine Poisson structures we rely on
Gelfand-Fuks cohomology. Since the handling of this cohomology
theory is not obvious, we derive, in the Appendix, an elementary,
``hands-on'' computation of the two first Gelfand-Fuks cohomological
groups of $\VectS$.


\section{Hamiltonian and bi-Hamiltonian
manifolds}\label{sec:Poisson}

In this section, we recall definitions and well-known results on
finite dimensional smooth Poisson manifolds.

\subsection{Poisson manifolds}
\label{subsec:Poisson_Manifolds}

\begin{definition}
A \emph{symplectic manifold} is a pair $(M,\omega)$, where $M$ is a
manifold and $\omega$ is a closed nondegenerate $2$-form on $M$,
that is $\rd\omega=0$ and for each $m \in M$, $\omega_{m}$ is a non
degenerate bilinear skew-symmetric map of $T_{m}M$.
\end{definition}

Since a symplectic form $\omega$ is nondegenerate, it induces an
isomorphism
\begin{equation}\label{equ:symplectic_iso}
    TM  \to T^{*}M,\quad X  \mapsto  i_{X}\,\omega ,
\end{equation}
defined via $i_{X}\,\omega(Y) = \omega(X,Y)$. For example, this
allows to define the \emph{symplectic gradient} $X_{f}$ of a
function $f$ by the relation $i_{X_{f}}\,\omega = -\rd f$. The
inverse of this isomorphism \eqref{equ:symplectic_iso} defines a
skew-symmetric bilinear form $P$ on the cotangent space $T^{*}M$.
This bilinear form $P$ induces itself a bilinear mapping on $\CM$,
the space of smooth functions $f: M \to \mathbb{R}$, given by
\begin{equation}\label{equ:symplectic_bracket}
    \poisson{f}{g} = P(\rd f,\rd g) = \omega(X_{f},X_{g}),
    \quad f,g \in \CM ,
\end{equation}
and called the \emph{Poisson bracket} of the functions $f$ and $g$.

The observation that a bracket like \eqref{equ:symplectic_bracket}
could be introduced on $C^\infty(M)$ for a smooth manifold $M$,
without the use of a symplectic form, leads to the general notion of
a \emph{Poisson structure} \cite{Lic77}.

\medskip

\begin{definition}
A \emph{Poisson (or Hamiltonian\footnote{The expression
\emph{Hamiltonian manifold} is often used for the generalization of
Poisson structure in the case of infinite dimension manifolds.})
structure} on a $C^{\infty}$ manifold $M$ is a skew-symmetric
bilinear mapping $(f,g)\mapsto \poisson{f}{g}$ on the space $\CM$,
which satisfies the \emph{Jacobi identity}
\begin{equation}\label{equ:Jacobi_identity}
    \poisson{\poisson{f}{g}}{h} + \poisson{\poisson{g}{h}}{f} +
    \poisson{\poisson{h}{f}}{g} = 0,
\end{equation}
as well as the \emph{Leibnitz identity}
\begin{equation}\label{equ:Leibnitz_identity}
    \poisson{f}{gh} = \poisson{f}{g}h + g\poisson{f}{h}.
\end{equation}
\end{definition}

\medskip

When the Poisson structure is induced by a symplectic structure
$\omega$, the \emph{Leibnitz identity} is a direct consequence of
\eqref{equ:symplectic_bracket}, whereas the \emph{Jacobi
identity}~\eqref{equ:Jacobi_identity} corresponds to the condition
$\rd\omega = 0$ satisfied by the symplectic form $\omega$. In the
general case, the fact that the mapping $g\mapsto\poisson{f}{g}$
satisfies \eqref{equ:Leibnitz_identity} means that it is a
\emph{derivation} of $\CM$.

Each derivation on $C^\infty(M)$ corresponds to a smooth vector
field, that is, to each $f \in \CM$ is associated a vector field
$X_{f}:M \to TM$, called the \emph{Hamiltonian vector field} of $f$,
such that
\begin{equation}\label{equ:def_Hamiltonian}
    \poisson{f}{g} = X_{f} \cdot g = L_{X_f}\, g,
\end{equation}
where $L_{X_f}\, g$ is the \emph{Lie derivative} of $g$ along $X_f$.

Jost \cite{Jos64} pointed out that, just like a derivation on
$C^\infty(M)$ corresponds to a vector field, a bilinear bracket
$\poisson{f}{g}$ satisfying the Leibnitz
rule~\eqref{equ:Leibnitz_identity} corresponds to a field of
bivectors. That is, there exists a $C^{\infty}$ tensor field
$P\in\Gamma(\bigwedge^{2}TM)$, called the \emph{Poisson bivector} of
$(M,\poisson{\cdot}{\cdot})$, such that
\begin{equation}\label{equ:Poisson_bracket}
    \poisson{f}{g}=P(\rd f,\rd g).
\end{equation}
for all $f,g \in \CM$.

\begin{proposition}
A bivector field $P\in\Gamma(\bigwedge^{2}TM)$ is the \emph{Poisson
bivector} of a Poisson structure on $M$ if and only if one of the
following equivalent conditions holds:
\begin{enumerate}
  \item $[P,P]=0$, where $[\; , \,]$ is the \emph{Schouten-Nijenhuis bracket}\footnote{The Schouten-Nijenhuis bracket is an extension of the Lie bracket of vector fields to skew-symmetric
multivector fields, see~\cite{Vai94}.},

  \item The bracket $\poisson{f}{g}=P(\rd f,\rd g)$ satisfies the Jacobi
  identity,

  \item $[X_{f},X_{g}] = X_{\poisson{f}{g}}$, for all $f,g \in \CM$.
\end{enumerate}
\end{proposition}

\begin{proof}
By definition of the Schouten-Nijenhuis bracket \cite{Vai94}, we
have
\begin{align*}
    -\frac{1}{2}\, [P,P] (\rd f,\rd g,\rd h)
        & = \circlearrowleft P(\rd Q(\rd f,\rd g),\rd h) \\
        & = \poisson{\poisson{f}{g}}{h} + \poisson{\poisson{g}{h}}{f} +
        \poisson{\poisson{h}{f}}{g} \\
        & = X_{\poisson{f}{g}}\cdot h - X_{f}\cdot X_{g} \cdot h + X_{g}\cdot X_{f} \cdot h
\end{align*}
for all $f,g,h \in \CM$ where $\circlearrowleft$ indicates the sum
over circular permutations of $f,g,h$. Hence, all these expressions
vanish together.
\end{proof}

\begin{remark}
The notion of a Poisson manifold is more general than that of a
symplectic manifold. Symplectic structures correspond to
nondegenerate Poisson structure. In that case, the Poisson bracket
satisfies the additional property that $\poisson{f}{g}=0$ for all $g
\in C^\infty(M)$ only if $f \in C^\infty(M)$ is a constant, whereas
for Poisson manifolds such non-constant functions $f$ might exist,
in which case they are called \emph{Casimir functions}. Such
functions are constants of motion for all vector fields $X_g$ where
$g\in \CM$.
\end{remark}

On a Poisson manifold $(M,P)$, a vector field $X: M \to TM$ is said
to be \emph{Hamiltonian} if there exists a function $f$ such that
$X=X_{f}$. On a symplectic manifold $(M,\omega)$, a necessary
condition for a vector field $X$ to be Hamiltonian is that
\begin{equation*}
    L_{X}\omega=0 .
\end{equation*}
A similar criterion exists for a Poisson manifold $(M,P)$ (see
\cite{Vai94}). A necessary condition for a vector field $X$ to be
\emph{Hamiltonian} is
\begin{equation*}
    L_{X}P = 0.
\end{equation*}

\subsection{Integrability}
\label{subsec:Integrability}

An \emph{integrable system} on a symplectic manifold $M$ of
dimension $2n$ is a set of $n$ functionally
independent\footnote{This means that the corresponding Hamiltonian
vector fields $X_{f_{1}}, \dotsc ,X_{f_{n}}$ are independent on an
open dense subset of $M$.} $f_{1}, \dotsc , f_{n}$ which are
\emph{in involution}, i.e. such that
\begin{equation*}
    \forall j,k \qquad \poisson{f_{j}}{f_{_{k}}} = 0 .
\end{equation*}
A Hamiltonian vector field $X_{H}$ is said to be \emph{(completely)
integrable} if the Hamiltonian function $H$ belongs to an integrable
system. In other words, $X_{H}$ is integrable if there exists $n$
first integrals\footnote{A first integral is a function which is
constant on the trajectories of the vector field.} of $X_{H}$,
$f_{1} = H, f_{2}, \dotsc , f_{n}$ which commute together.

\begin{remark}
At any point $x$ where the functions $f_{1}, \dotsc , f_{n}$ are
functionally independent, the Hamiltonian vector fields $X_{f_{1}},
\dotsc ,X_{f_{n}}$ generate a \emph{maximal isotropic} subspace
$L_{x}$ of $T_{x}M$. When $x$ varies, the subspaces generate what
one calls a \emph{Lagrangian distribution}; that is a sub-bundle $L$
of $TM$ whose fibers are maximal isotropic subspaces. In our case,
this distribution is integrable (in the sense of Frobenius). The
leaves of $L$ are defined by the equations
\begin{equation*}
    f_{1} = \const, \dotsc , f_{n} = \const.
\end{equation*}
A Lagrangian distribution which is integrable (in the sense of
Frobenius) is called a \emph{real polarization} and is a key notion
in \emph{Geometric Quantization}.
\end{remark}

In the study of dynamical systems, the importance of integrable
Hamiltonian vector fields is emphasized by the
\emph{Arnold-Liouville theorem} \cite{Arn97} which asserts that each
compact leaf is actually diffeomorphic to an $n$-dimensional torus
\begin{equation*}
    T^{n} = \set{(\varphi^{1}, \dotsc ,\varphi^{n});\quad \varphi^{k} \in
    \mathbb{R}/2\pi\mathbb{Z}},
\end{equation*}
on which the flow of $X_{H}$ defines a linear quasi-periodic motion,
i.e. that in angular coordinates $(\varphi^{1}, \dotsc
,\varphi^{n})$
\begin{equation*}
    \dd{\varphi^{k}}{t} = \omega^{k}, \quad k=0 ,\dotsc , n,
\end{equation*}
where $(\omega^{1}, \dotsc , \omega^{n})$ is a constant vector.

\begin{remark}
In the case of a Poisson manifold, it can be confusing to define an
integrable system. However, we can use the symplectic definition on
each symplectic leaves of the Poisson manifold.
\end{remark}

\subsection{Bi-Hamiltonian manifolds}
\label{subsec:Bihamiltonian_manifolds}

Two Poisson brackets $\poisson{\;}{}_{P}$ and $\poisson{\;}{}_{Q}$
are \emph{compatible} if any linear combination
\begin{equation*}
\poisson{f}{g}_{\lambda,\,\mu} =
\lambda\poisson{f}{g}_{P}+\mu\poisson{f}{g}_{Q}, \qquad
\lambda,\mu\in\mathbb{R},
\end{equation*}
is also a Poisson bracket. A \emph{bi-Hamiltonian manifold}
$(M,P,Q$) is a manifold equipped with two Poisson structures $P$ and
$Q$ which are compatible.

\begin{proposition}\label{prop:compatibility_condition}
Let $P$ and $Q$ be two Poisson structures on $M$. Then $P$ and $Q$
are compatible if and only if one of the following equivalent
conditions holds:
\begin{enumerate}
  \item $[P,Q] = 0$, where $[\; , \,]$ is the Schouten-Nijenhuis
  bracket,
  \item $\circlearrowleft  \poisson{\poisson{g}{h}_{P}}{f}_{Q} + \poisson{\poisson{g}{h}_{Q}}{f}_{P} = 0$, where $\circlearrowleft$
  is the sum over circular permutations of $f,g,h$,
  \item $[X^{P}_{f},X^{Q}_{g}] + [X^{Q}_{f},X^{P}_{g}] = X^{P}_{\poisson{f}{g}_{Q}} + X^{Q}_{\poisson{f}{g}_{P}}$, for all $f,g \in
  \CM$.
\end{enumerate}
\end{proposition}

\begin{proof}
By definition of the Schouten-Nijenhuis bracket \cite{Vai94}, we
have
\begin{align*}
     - \, [P,Q] (\rd f,\rd g,\rd h)
        & = \circlearrowleft P(\rd Q(\rd f,\rd g),\rd h) + Q(\rd P(\rd f,\rd g),\rd h) \\
        & = \circlearrowleft  \poisson{\poisson{g}{h}_{P}}{f}_{Q} +
        \poisson{\poisson{g}{h}_{Q}}{f}_{P} \\
        & = - [X^{P}_{f},X^{Q}_{g}]\cdot h - [X^{Q}_{f},X^{P}_{g}]\cdot h\\
        & + X^{P}_{\poisson{f}{g}_{Q}}\cdot h + X^{Q}_{\poisson{f}{g}_{P}}\cdot h
\end{align*}
for all $f,g,h \in \CM$. Hence, all these expressions vanish
together.
\end{proof}

\subsection{Lenard recursion relations}
\label{subsec:Lenard}

On a bi-Hamiltonian manifold $M$, equipped with two compatible
Poisson structures $P$ and $Q$, we say that a vector field $X$ is
(formally) \emph{integrable}\footnote{This terminology is used for
evolution equations in infinite dimension.} or \emph{bi-Hamiltonian}
if it is Hamiltonian for both structures. The reason for this
terminology is that for such a vector field, there exists under
certain conditions a hierarchy of first integrals in involution that
may lead in certain case to complete integrability, in the sense of
Liouville. A useful concept for obtaining such a hierarchy of first
integrals is the so called \emph{Lenard scheme} \cite{McK93}.

\begin{definition}
On a manifold $M$ equipped with two Poisson structures $P$ and $Q$,
we say that a sequence $(H_{k})_{k\in \mathbb{N}^{*}}$ of smooth
functions satisfy the \emph{Lenard recursion relation} if
\begin{equation}\label{equ:Lenard_recursion_scheme}
    P \, dH_{k} = Q\, dH_{k+1},
\end{equation}
for all $k\in \mathbb{N}^{*}$.
\end{definition}

\begin{proposition}\label{prop:Lenard}
Let $P$ and $Q$ be Poisson structures on a manifold $M$ and let
$(H_{k})_{k\in \mathbb{N}^{*}}$ be a sequence of smooth functions on
$M$ that satisfy the Lenard recursion relation. Then the functions,
$H_{k}$, are pairwise in involution with respect to both brackets
$P$ and $Q$.
\end{proposition}

\begin{proof}
Using skew-symmetry of $P$ and $Q$ and relation
\eqref{equ:Lenard_recursion_scheme}, we get
\begin{equation*}
    P(dH_{k},dH_{k+p}) = Q(dH_{k+1},dH_{k+p}) =
    P(dH_{k+1},dH_{k+p-1}),
\end{equation*}
for all $k,p \in \mathbb{N}^{*}$. From which we deduce, by induction
on $p$, that
\begin{equation*}
    \poisson{H_{k}}{H_{k+p}}_{P} = 0,
\end{equation*}
for all $k,p \in \mathbb{N}^{*}$. It is then an immediate
consequence that
\begin{equation*}
    \poisson{H_{k}}{H_{l}}_{Q} = 0,
\end{equation*}
for all $k,l \in \mathbb{N}^{*}$.
\end{proof}

\begin{remark}
Notice that in the proof of proposition~\ref{prop:Lenard}, the
compatibility of $P$ and $Q$ is not needed.
\end{remark}

Suppose now that $(M,P,Q)$ is a bi-Hamiltonian manifold and that at
least one of the two Poisson brackets, say $Q$ is \emph{invertible}.
In that case, we can define a $(1,1)$-tensor field
\begin{equation*}
    R = PQ^{-1},
\end{equation*}
which is called the \emph{recursion operator} of the bi-Hamiltonian
structure. It has been shown \cite{KSM90,KSM96} that, as a
consequence of the compatibility of $P$ and $Q$, the \emph{Nijenhuis
torsion} of $R$, defined by
\begin{equation*}
    T(R)(X,Y) = [RX,RY] - R\big( [RX,Y] + [X,RY] \big) + R^{2}[X,Y]
\end{equation*}
vanishes. In this situation, the family of Hamiltonians
\begin{equation*}
    H_{k} = \frac{1}{k} \tr R^{k}, \qquad (k \in \mathbb{N}^{*}),
\end{equation*}
satisfy the Lenard recursion relation
\eqref{equ:Lenard_recursion_scheme}. Indeed, this results from the
fact that
\begin{equation*}
    L_{X} \, \tr (T) = \tr (L_{X} \, T)
\end{equation*}
for every vector field $X$ and every $(1,1)$-tensor field $T$ on M and
that the vanishing of the Nijenhuis torsion of $R$ can be rewritten
as
\begin{equation*}
    L_{RX} \, R = R \, L_{X}\, R
\end{equation*}
for all vector field $X$.

\begin{remark}
This construction has to be compared with \emph{Lax isospectral
equation} associated to an evolution equation
\begin{equation}\label{equ:evolution_equation}
    \dd{u}{t} = F(u).
\end{equation}
The idea is to associate to equation~\eqref{equ:evolution_equation},
a pair of matrices (or operators in the infinite dimensional case)
$(L,B)$, called a \emph{Lax pair}, whose coefficients are functions
of $u$ and in such a way that when $u(t)$ varies according to
\eqref{equ:evolution_equation}, $L(t) = L(u(t))$ varies according to
\begin{equation*}
    \dd{L}{t} = [L,B].
\end{equation*}
This equation has been formulated in \cite {Lax68} in order to
obtain a hierarchy of first integrals of the evolution equation as
eigenvalues or traces of the operator $L$. This analogy between $R$
and $L$ is not casual and has been studied in \cite{KSM96}. Many
evolution equations which admit a Lax pair appear to be also
bi-Hamiltonian systems generated by a recursion operator
$R=PQ^{-1}$.
\end{remark}

\medskip

In practice, we may be confronted to the following problem. We start
with an evolution equation represented by a vector field $X$ on a
manifold $M$. We find two compatible Poisson structures $P$ and $Q$
on $M$ which makes $X$ a bi-Hamiltonian vector field. But $P$ and
$Q$ are \emph{both non-invertible}. In that case, it is however
still possible to find a Lenard hierarchy if the following algorithm
works.

\textit{Step 1:} Let $H_{1}$ the Hamiltonian of $X$ for the Poisson
structure $P$ and let $X_{1}=X$. The vector field $X_{1}$ is
Hamiltonian for the Poisson structure $Q$ by assumption, this
defines Hamiltonian function $H_{2}$. We define $X_{2}$ to be the
Hamiltonian vector field generated by $H_{2}$ for the Poisson
structure $P$.

\textit{Step 2:} Inductively, having defined Hamiltonian function
$H_{k}$ and letting $X_{k}$ be the Hamiltonian vector field
generated by $H_{k}$ for the Poisson structure $P$, we check if
$X_{k}$ is Hamiltonian for the Poisson structure $Q$. If the answer
is yes, then we define $H_{k+1}$ to be the Hamiltonian of $X_{k}$
for the Poisson structure $Q$.


\section{Poisson structures on the dual of a Lie algebra}
\label{sec:Lie_Poisson}

\subsection{Lie-Poisson structure}
\label{subsec:Lie_Poisson}

The fundamental example of a non-symplectic Poisson structure is the
\emph{Lie-Poisson structure} on the dual $\g^{*}$ of a Lie algebra
$\g$.

\begin{definition}
On the dual space $\g^{*}$ of a Lie algebra $\g$ of a Lie group $G$,
there is a Poisson structure defined by
\begin{equation}\label{equ:LiePoisson}
    \poisson{f}{g}(m) = m([\rd_{m}f,\rd_{m}g])
\end{equation}
for $m\in\g^{*}$ and $f,g \in C^{\infty}(\g^{*})$, called the
\emph{canonical Lie-Poisson structure}\footnote{Here, $\rd_{m}f$,
the differential of a function $f \in C^\infty(\g^{*})$ at $m \in
\g^{*}$ is to be understood as an element of the Lie algebra $\g$}.
\end{definition}

\begin{remark}
The canonical Lie-Poisson structure has the remarkable property to
be \emph{linear}, that is the bracket of two linear functionals is
itself a linear functional. Given a basis of $\g$, the
components\footnote{In what follows, the convention for lower or
upper indices may be confusing since we shall deal with tensors on
both $\g$ and $\g^{*}$. Therefore, we emphasize that the convention
we use in this paper is the following: upper-indices correspond to
contravariant tensors on $\g$ and therefore covariant tensors on
$\g^{*}$ whereas lower indices correspond to covariant tensors on
$\g$ and therefore contravariant tensors on $\g^{*}$.} of the
Poisson bivector $W$ associated to \eqref{equ:LiePoisson} are
\begin{equation}\label{equ:Poisson_Lie_bivector}
    P_{ij} = c_{ij}^{k}\,x_{k},
\end{equation}
where $c_{ij}^{k}$ are the \emph{structure component} of the Lie
algebra $\g$.
\end{remark}

\subsection{Modified Lie-Poisson structures}
\label{subsec:Modified_Lie_Poisson}

Under the general name of \emph{modified Lie-Poisson structures}, we
mean an affine\footnote{A Poisson structure on a linear space is
\emph{affine} if the bracket of two linear functionals is an affine
functional.} perturbation of the canonical Lie-Poisson structure on
$\g^{*}$. In other words, it is represented by a bivector
\begin{equation*}
P + Q,
\end{equation*}
where $P$ is the canonical Poisson bivector defined by
\eqref{equ:Poisson_Lie_bivector} and $Q = (Q_{ij})$ is a constant
bivector on $\g^{*}$. Such a $Q\in\bigwedge^{2}\g^{*}$ is itself a
Poisson bivector. Indeed the Schouten-Nijenhuis bracket
\begin{equation*}
[Q , Q]=0,
\end{equation*}
since $Q$ is a constant tensor field on $\g^{*}$.

The fact that $P+Q$ is a Poisson bivector, or equivalently that $Q$
is compatible with the canonical Lie-Poisson structure, is expressed
using proposition~\ref{prop:compatibility_condition}, by the
condition
\begin{equation}\label{equ:cocycle_condition}
    Q([u,v],w) + Q([v,w],u) + Q([w,u],v)=0,
\end{equation}
for all $u,v,w \in \g$.

\subsection{Lie algebra cohomology}
\label{subsec:Lie_algebra_cohomology}

On a Lie group $G$, a left-invariant\footnote{In this section, we
deal with left-invariant forms but, of course, everything we say may
be applied equally to right-invariant forms up to a sign in the
definition of the coboundary operator.} $p$-form $\omega$ is
completely defined by its value at the unit element $e$, and hence
by an element of $\bigwedge^{p}\g^{*}$. In other words, there is a
natural isomorphism between the space of left-invariant $p$-forms on
$G$ and $\bigwedge^{p}\g^{*}$. Moreover, since the exterior
differential $\rd$ commutes with left translations, it induces a
linear operator $\partial:\bigwedge^{p}\g^{*} \to
\bigwedge^{p+1}\g^{*}$ defined by
\begin{equation}\label{equ:coboundary_operator}
    \partial \gamma(u_{0}, \dotsc ,u_{p}) =
    \sum_{i<j}(-1)^{i+j}\gamma([u_{i},u_{j}], u_{0}, \dotsc , \widehat{u_{i}}, \dotsc, \widehat{u_{j}}, \dotsc ,u_{p}) ,
\end{equation}
where the hat means that the corresponding element should not appear
in the list. $\gamma$ is said to be a \emph{cocycle} if $\partial
\gamma =0$. It is a \emph{coboundary} if is of the form $\gamma =
\partial\mu$ for some cochain $\mu$ in dimension $p-1$. Every
coboundary is a cocycle: that is $\partial \circ \partial = 0$.

\begin{example}
For every $\gamma\in\bigwedge^{0}\g^{*}=\mathbb{R}$, we have
$\partial\gamma=0$. For $\gamma\in\bigwedge^{1}\g^{*}=\g^{*}$, we
have
\begin{equation*}
    \partial\gamma\,(u,v)=-\gamma([u,v]),
\end{equation*}
where $u,v \in \g$. For $\gamma\in\bigwedge^{2}\g^{*}$, we have
\begin{equation*}
    \partial\gamma\,(u,v,w)= - \gamma([u,v],w) - \gamma([v,w],u) -
    \gamma([w,u],v),
\end{equation*}
where $u,v,w \in \g$.
\end{example}

The kernel $Z^{p}(\g)$ of $\partial: \bigwedge^{p}(\g^{*}) \to
\bigwedge^{p+1}(\g^{*})$ is the space of \emph{$p$-cocycles} and the
range $B^{p}(\g)$ of $\partial: \bigwedge^{p-1}(\g^{*}) \to
\bigwedge^{p}(\g^{*})$ is the spaces of \emph{$p$-coboundaries}. The
quotient space $H^{p}_{CE}(\g)= Z^{p}(\g)/B^{p}(\g)$ is the $p$-th
\emph{Lie algebra cohomology} or \emph{Chevaley-Eilenberg cohomology
group} of $\g$. Notice that in general the Lie algebra cohomology is
different from the de Rham cohomology $H^{p}_{DR}$. For example,
$H^{1}_{DR}(\mathbb{R}) = \mathbb{R}$ but $H^{1}_{CE}(\mathbb{R})
=0$.

\begin{remark}
Each  $2$-cocycle $\gamma$ defines a modified Lie-Poisson structure
on $\g^{*}$. The compatibility
condition~\eqref{equ:cocycle_condition} can be recast as $\partial
\gamma=0$. Notice that the Hamiltonian vector field $X_{f}$ of a
function $f\in C^{\infty}(\g^{*})$ computed with respect to the
Poisson structure defined by the $2$-cocycle $\gamma$ is
\begin{equation}\label{equ:Hamiltonian_for_CS}
    X_{f}(m) = \gamma (\rd_mf, \cdot).
\end{equation}
\end{remark}

\begin{example}
A special case of modified Lie-Poisson structure is given by a
$2$-cocycle $\gamma$ which is a coboundary. If $\gamma = \partial
m_{0}$ for some $m_{0}\in \g^{*}$, the expression
\begin{equation*}
    \poisson{f}{g}_{0} (m) = m_{0}([\rd_{m}f,\rd_{m}g])
\end{equation*}
looks like if the Lie-Poisson bracket had been ``frozen'' at a point
$m_{0} \in \g^{*}$ and for this reason some authors call it a
\emph{freezing} structure.
\end{example}


\section{Bi-Hamiltonian vector fields on $\VectSstar$}
\label{sec:VectS}

\subsection{The Lie algebra $\VectS$}
\label{subsec:VectS}

The group $\mathfrak{D}$ of smooth orientation-preserving
diffeomorphisms of the circle $\mathbb{S}^1$ is endowed with a
smooth manifold structure based on the \emph{Fr\'{e}chet space}
$C^{\infty}(\Ucircle)$. The composition and the inverse are both
smooth maps $\mathfrak{D}\times \mathfrak{D} \to \mathfrak{D}$,
respectively $\mathfrak{D} \to \mathfrak{D}$, so that $\mathfrak{D}$
is a Lie group \cite{Mil84}. Its Lie algebra $\mathfrak{g}$ is the
space $\VectS$ of smooth vector fields on $\Ucircle$, which is
itself isomorphic to the space $C^{\infty}(\Ucircle)$ of periodic
functions. The Lie bracket\footnote{It corresponds to the Lie
bracket of right-invariant vector fields on the group.} on
$\mathfrak{g} = \VectS$ is given by
\begin{equation*}
    [u,v] = uv_{x} - u_{x}v .
\end{equation*}

\begin{lemma}\label{lem:commutator_algebra}
The Lie algebra $\VectS$ is equal to its commutator algebra. That is
\begin{equation*}
    \left[\VectS,\VectS \right] = \VectS.
\end{equation*}
\end{lemma}

\begin{proof}
Any real periodic function $u$ on can be written uniquely as the sum
\begin{equation*}
    u = w + c
\end{equation*}
where $w$ is periodic function of total integral zero and $c$ is a
constant. To be of total integral zero is the necessary and
sufficient condition for a periodic function $w$ to have a periodic
primitive $W$. Hence we have $[1, W] = w$. Moreover, since
$[\cos,\sin] = 1$, we have proved that every periodic function $u$
can be written as the sum of two commutators.
\end{proof}

\subsection{The regular dual $\VectSstar$}
\label{subsec:VectS_dual}

Since the topological dual of the Fr\'{e}chet space $\VectS$ is too big
and not tractable for our purpose, being isomorphic to the space of
distributions on the circle, we restrict our attention in the
following to the \emph{regular dual} $\mathfrak{g}^{*}$, the
subspace of $\VectS^{*}$ defined by linear functionals of the form
\begin{equation*}
    u \mapsto \int_{\Ucircle}mu\, dx
\end{equation*}
for some function $m \in C^{\infty}(\Ucircle)$. The regular dual
$\mathfrak{g}^{*}$ is therefore isomorphic to $C^{\infty}(\Ucircle)$
by means of the $L^2$ inner product\footnote{In the sequel, we use
the notation $u,v,\dotsc$ for elements of $\mathfrak{g}$ and
$m,n,\dots$ for elements of $\mathfrak{g}^{*}$ to distinguish them,
although they all belong to $C^{\infty}(\Ucircle)$.}
\begin{equation*}
    <u , v > = \int_{\Ucircle} uv \, dx .
\end{equation*}
With these definitions, the \emph{coadjoint action}\footnote{The
coadjoint action of a Lie algebra $\g$ on its dual is defined as
\begin{equation*}
    (ad_{u}\, m , v) = -(m, ad_{u}\, v) = -(m, [u,v]),
\end{equation*}
where $u,v \in \g$, $m\in \g^{*}$ and the pairing is the standard
one between $\g$ and $\g^{*}$.} of the Lie algebra $\VectS$ on the
regular dual $\VectSstar$ is given by
\begin{equation*}
    ad^{*}_{u} \, m = mu_{x} + (mu)_{x} = 2mu_{x} + m_{x}u.
\end{equation*}

Let $F$ be a smooth real valued function on $C^{\infty}(\Ucircle)$.
Its \emph{Fr\'{e}chet} derivative $\rd F(m)$ is a linear functional on
$C^{\infty}(\Ucircle)$. We say that $F$ is a \emph{regular function}
if there exists a smooth map $\delta F : C^{\infty}(\Ucircle) \to
C^{\infty}(\Ucircle)$ such that
\begin{equation*}
    \rd F(m)\,M = \int_{\Ucircle} M\cdot\delta F (m)
 \, dx,\qquad m, M \in C^{\infty}(\Ucircle).
\end{equation*}
That is, the Fr\'{e}chet derivative $\rd F(m)$ belongs to the regular
dual $\mathfrak{g}^{*}$ and the mapping $m \mapsto \delta F(m)$ is
smooth. The map $\delta F$ is a vector field on
$C^{\infty}(\Ucircle)$, called the \emph{gradient} of $F$ for the
$L^{2}$-metric. In other words, a regular function is a smooth
function on $C^{\infty}(\Ucircle)$ which has a smooth $L^{2}$
gradient.

\medskip

\begin{example}
Typical examples of \emph{regular functions} on the space
$C^{\infty}(\Ucircle)$ are \emph{linear functionals}
\begin{equation*}
    F(m) =  \int_{\Ucircle} um  \, dx ,
\end{equation*}
where $u \in C^{\infty}(\Ucircle)$. In that case, $\delta F(m)=u$.
Other examples are \emph{nonlinear polynomial functionals}
\begin{equation*}
    F(m) =  \int_{\Ucircle} Q(m) \, dx ,
\end{equation*}
where $Q$ is a polynomial in derivatives of $m$ up to a certain
order $r$. In that case,
\begin{equation*}
    \delta F(m)= \sum_{k=0}^{r} (-1)^{k} \frac{d^{k}}{dx^{k}} \left( \frac{\partial Q}{\partial X_{k}}
    (m)\right).
\end{equation*}
Notice that the smooth function $F_\theta: C^{\infty}(\Ucircle) \to
\mathbb{R}$ defined by $F_\theta(m)=m(\theta)$ for some fixed
$\theta \in \mathbb{S}^1$ is not regular since $\rd F_\theta$ is the
Dirac measure at $\theta$.
\end{example}

\medskip

A smooth vector field $X$ on $\mathfrak{g}^{*}$ is called a
\emph{gradient} if there exists a \emph{regular function} $F$ on
$\mathfrak{g}^{*}$ such that $X(m) = \delta F(m)$ for all $m \in
\mathfrak{g}^{*}$. Observe that if $F$ is a smooth real valued
function on $C^{\infty}(\Ucircle)$ then its second Fr\'{e}chet
derivative is symmetric \cite{Ham82}, that is,
\begin{equation*}
    \rd^{2}F(m)(M,N) = \rd^{2}F(m)(N,M), \qquad m,M,N \in
    C^{\infty}(\Ucircle) .
\end{equation*}

For a regular function, this property can be rewritten as
\begin{equation}\label{equ:symmetric_condition}
    \int_{\Ucircle} \Big(\delta F^{\prime}(m)M \Big)N  \, dx =
\int_{\Ucircle} \Big( \delta F^{\prime}(m)N \Big)M  \,
    dx ,
\end{equation}
for all $m,M,N \in C^{\infty}(\Ucircle)$. That is, the linear
operator $\delta F^{\prime}(m)$ is symmetric for the $L^{2}$-inner
product on $C^{\infty}(\Ucircle)$ for each $m \in
C^{\infty}(\Ucircle)$. Conversely, a smooth vector field $X$ on
$\mathfrak{g}^{*}$ whose Fr\'{e}chet derivative $X^{\prime}(m)$ is a
symmetric linear operator is the gradient of the function
\begin{equation}\label{equ:primitive}
    F(m) = \int_{0}^{1} <X(tm),m> \, dt .
\end{equation}
This can be checked directly, using the symmetry of $X^{\prime}(m)$
and an integration by part. We will resume this fact in the
following lemma.

\begin{lemma}\label{lem:gradient_condition}
On the Fr\'{e}chet space $C^{\infty}(\Ucircle)$ equipped with the (weak)
$L^{2}$ inner product, a necessary and sufficient condition for a
smooth vector field $X$ to be a \emph{gradient} is that its Fr\'{e}chet
derivative $X^{\prime}(m)$ is a symmetric linear operator.
\end{lemma}

\subsection{Hamiltonian structures on $\VectSstar$}
\label{subsec:VectS_Hamiltonian_structure}

To define a \emph{Poisson bracket} on the space of \emph{regular
functions} on $\mathfrak{g}^{*}$, we consider a one-parameter family
of linear operators $P_{m}$ ($m\in\CS$) and set
\begin{equation}\label{equ:regular_Poisson_bracket}
    \poisson{F}{G}(m) = \int_{\Ucircle} \delta F(m)
\, P_{m} \, \delta G(m)  \,
    dx .
\end{equation}
The operators $P_{m}$ must satisfy certain conditions in order for
\eqref{equ:regular_Poisson_bracket} to be a valid Poisson structure
on the regular dual $\mathfrak{g}^{*}$.

\begin{definition}
A family of linear operators $P_{m}$ on $\mathfrak{g}^{*}$ define a
Poisson structure on $\mathfrak{g}^{*}$ if
~\eqref{equ:regular_Poisson_bracket} satisfies
\begin{enumerate}
    \item $\poisson{F}{G}$ is regular if $F$ and $G$ are regular,
    \item $\poisson{G}{F} = - \poisson{F}{G}$,
    \item $\poisson{\poisson{F}{G}}{h} + \poisson{\poisson{G}{H}}{F}
    + \poisson{\poisson{H}{F}}{G} = 0$.
\end{enumerate}
\end{definition}

\medskip

Notice that the second condition above simply means that $P_{m}$ is
a skew-symmetric operator for each $m$.

\medskip

\begin{example}
The canonical Lie-Poisson structure on $\mathfrak{g}^{*}$ given by
\begin{equation*}
    \poisson{F}{G}(m) = m \left( \brackets{\delta F}{\delta G} \right)
    = \int_{\Ucircle} \delta F (m) \left( mD + Dm \right) \delta
    G(m)\,dx
\end{equation*}
is represented by the one-parameter family of skew-symmetric
operators
\begin{equation}\label{equ:Lie_Poisson_operator}
    P_{m} = mD + Dm
\end{equation}
where $D=\partial_x$. It can be checked that all the three required
properties are satisfied. In particular, we have
\begin{equation*}
    \delta \poisson{F}{G} = \delta F^{\prime} (P_{m} \delta G) -
\delta G^{\prime} (P_{m} \delta
    F) + \delta F \, D\delta G - \delta G \, D\delta F .
\end{equation*}
\end{example}

\medskip

\begin{definition}
The \emph{Hamiltonian} of a \emph{regular} function $F$, for a
Poisson structure defined by $P$ is defined as the vector field
\begin{equation*}
    X_{F}(m) = P \, \delta F(m).
\end{equation*}
\end{definition}

\medskip

\begin{proposition}\label{prop:Hamiltonian_criterium}
A necessary condition for a smooth vector field $X$ on
$\mathfrak{g}^{*}$ to be \emph{Hamiltonian} with respect to the
Poisson structure defined by a \emph{constant} linear operator $Q$
is the symmetry of the operator $X^{\prime}(m)Q$ for each $m\in
\mathfrak{g}^{*}$.
\end{proposition}

\begin{proof}
If $X$ is Hamiltonian, we can find a regular function $F$ such that
\begin{equation*}
    X(m) = Q \delta F(m) .
\end{equation*}
Moreover, since $Q$ is a constant linear operator, we have
\begin{equation*}
    X^{\prime}(m) = Q \delta F^{\prime}(m),
\end{equation*}
and therefore, we get
\begin{equation*}
    X^{\prime}(m)Q = Q \delta F^{\prime}(m) Q ,
\end{equation*}
which is a symmetric operator since $Q$ is skew-symmetric and
$\delta F^{\prime}(m)$ is symmetric.
\end{proof}

\subsection{Hamiltonian vector fields generated by right-invariant metrics}
\label{subsec:right_invariant_metrics}

A right-invariant metric on the diffeomorphism group
$Diff(\Ucircle)$ is uniquely defined by its restriction to the
tangent space to the group at the unity, hence by a
\emph{non-degenerate continuous inner product} $\mathbf{a}$ on
$\VectS$. If this inner product $\mathbf{a}$ is \emph{local}, then
according to Peetre~\cite{Pee59}, there exists a linear differential
operator
\begin{equation}\label{equ:def_A}
    A = \sum_{j=0}^{N} a_{j} \frac{d^{j}}{dx^{j}}
\end{equation}
where $a_{j} \in C^{\infty}(\Ucircle)$ for $j=0,\dotsc , N$, such
that
\begin{equation*}
    \mathbf{a}(u,v) = \int_{\Ucircle} A(u)\, v \, dx = \int_{\Ucircle} A(v)\, u \, dx ,
\end{equation*}
for all $u,v \in \VectS$. The condition for $\mathbf{a}$ to be
non-degenerate is equivalent for $A$ to be a \emph{continuous linear
isomorphism} of $C^{\infty}(\Ucircle)$.

\begin{remark}
In the special case where $A$ has \emph{constant coefficients}, the
\emph{symmetry} is traduced by the fact that $A$ contains only even
derivatives and the \emph{non-degeneracy} by the fact that the
\emph{symbol} of $A$
\begin{equation*}
    s_{A}(\xi) = \re^{\ri x\xi} A(\re^{-\ri x\xi}) = \sum_{j=0}^{N} a_{2j} (-\ri \xi)^{2j},
\end{equation*}
has no root in $\mathbb{Z}$.
\end{remark}

The right-invariant metric on $\DiffS$ induced by a continuous,
linear, invertible operator $A$ gives rise to an \emph{Euler
equation}\footnote{The second order geodesic equation corresponding
to a one sided invariant metric on a Lie group can always be reduced
to a first order quadratic equation on the dual of the Lie algebra
of the group: the Euler equation (see \cite{AK98} or \cite{Kol04}).
The generality of this reduction was first revealed by Arnold
\cite{Arn66}.} on $\VectS^{*}$
\begin{equation}\label{equ:Euler_equation}
    \dd{m}{t} = 2mu_{x} + m_{x}u,
\end{equation}
where $m = Au$. This equation is Hamiltonian with respect to the
Lie-Poisson structure on $\VectS^{*}$ with Hamiltonian function on
$\VectS^{*}$ given by
\begin{equation*}
    H_{2}(m) = \frac{1}{2} \int_{\Ucircle} m u \, dx .
\end{equation*}
The corresponding Hamiltonian vector field $X_{A}$ is given by
\begin{equation*}
    X_{A}(m) = (mD + Dm)(A^{-1}m) = 2mu_{x} + um_{x} .
\end{equation*}

\begin{remark}
The family of operators
\begin{equation*}
A_k = 1-\frac{d^2}{dx^2}+ \dotsb +(-1)^k \frac{d^{2k}}{dx^{2k}},
\end{equation*}
corresponding respectively to the Sobolev $H^{k}$ inner product,
have been studied in \cite{CK02,CK03}. The \emph{Riemannian
exponential map} of the corresponding geodesic flow has been shown
to be a local diffeomorphism except for $k=0$. This later case
corresponds to the $L^{2}$ metric on $\DiffS$ and happens to be
\emph{singular}.
\end{remark}

\begin{remark}
A non-invertible inertia operator $A$ may induce in some cases, a
weak Riemannian metric on a \emph{homogenous space}. This is the way
to interpret Hunter-Saxton and Harry Dym equations as Euler
equations, see \cite{KM03}.
\end{remark}

The following theorem is a generalization of \cite[Theorem
3.7]{CK06}.

\begin{theorem}\label{thm:main}
The only continuous, linear, invertible operators
\begin{equation*}
A: \VectS \to \VectS^{*}
\end{equation*}
with \emph{constant coefficients}, whose corresponding Euler vector
field $X_{A}$ is bi-Hamiltonian relatively to some modified
Lie-Poisson structure are
\begin{equation*}
    A = aI + bD^{2},
\end{equation*}
where $a,b \in \mathbb{R}$ satisfy $a-bn^{2}\ne 0, \forall n \in
\mathbb{Z}$. The second Hamiltonian structure is induced by the
operator
\begin{equation*}
    Q = DA = aD + bD^{3},
\end{equation*}
where $D = d/dx$ and the Hamiltonian function is
\begin{equation*}
    H_{3}(m) = \frac{1}{2} \int_{\Ucircle} \big( au^{3} - bu(u_{x})^{2} \big) \,
    dx ,
\end{equation*}
where $m = Au$.
\end{theorem}

\begin{remark}
We insist on the fact that the proof we give applies for an operator
with \emph{constant coefficients}. It would be interesting to study
the case of an invertible, continuous linear operator whose
coefficients are \emph{not constant}. Are there such operator $A$
with bi-Hamiltonian Euler vector field $X_{A}$ relative to some
modified Lie-Poisson structure ? In that case, for which modified
Lie-Poisson structures $Q$ is there an Euler vector field $X_{A}$
which is bi-Hamiltonian relatively to $Q$~?
\end{remark}

\begin{proof}
The proof is essentially the same as the one given in \cite{CK06}. A
direct computation shows that
\begin{equation*}
    X_{A}(m) = (aD + bD^{3}) \, \delta H_{3}(m)
\end{equation*}
where
\begin{equation*}
    H_{3}(m) = \frac{1}{2} \int_{\Ucircle} \big( au^{3} - bu(u_{x})^{2} \big) \,
    dx ,
\end{equation*}
and
\begin{equation*}
    A = aI + bD^{2},
\end{equation*}
where $a,b \in \mathbb{R}$.

Each modified Lie-Poisson structure on $\VectSstar$ is given by a
\emph{local $2$-cocycle} of $\VectS$. According to
proposition~\ref{prop:second_cohomology_group} (see the Appendix),
such a cocycle is represented by a differential operator
\begin{equation}\label{equ:MLP_operator}
Q = m_{0}D + Dm_{0} + \beta D^{3}
\end{equation}
where $m_{0}\in \CS$ and $\beta \in \mathbb{R}$. We will now show
that there is no such cocycle for which $X_{A}$ is Hamiltonian if
the order of
\begin{equation*}
    A = \sum_{j=0}^{N} a_{2j} D^{2j}
\end{equation*}
is strictly greater than $2$.

By virtue of proposition~\ref{prop:Hamiltonian_criterium}, a
necessary condition for $X_{A}$ to be Hamiltonian with respect to
the cocycle represented by $Q$ is that
\begin{equation*}
    K(m) = X_{A}^{\prime}(m) Q
\end{equation*}
is a symmetric operator. We have
\begin{equation*}
    X_{A}^{\prime}(m) = 2u_{x}I + uD + 2m DA^{-1} + m_{x}A^{-1} ,
\end{equation*}
and in particular, for $m=1$,
\begin{equation*}
    X_{A}^{\prime} (1) = D + 2DA^{-1} .
\end{equation*}
Hence
\begin{equation*}
    K(1) = \big( D + 2DA^{-1} \big) \circ
    \big( m_{0}D + Dm_{0} \big) + \beta D^{4} ( 1 + 2A^{-1}) ,
\end{equation*}
whereas
\begin{equation*}
    K(1)^{*} = \big( m_{0}D + Dm_{0} \big) \circ \big( D + 2DA^{-1}
    \big) + \beta D^{4} ( 1 + 2A^{-1}) .
\end{equation*}
Therefore, letting $m_{0}^{\prime} = \dd{m_{0}}{x}$, we get
\begin{multline*}
    K(1) - K(1)^{*} = \big( m_{0}^{\prime} D + D m_{0}^{\prime} \big)
     + 2 \big( A^{-1}Dm_{0}D - Dm_{0}DA^{-1} \big) + \\
     + 2 \big( A^{-1}D^{2}m_{0} - m_{0}D^{2}A^{-1} \big) ,
\end{multline*}
and this operator vanishes if and only if
\begin{equation}\label{equ:null_criteria}
    A \big( K(1) - K(1)^{*} \big) A = 0 .
\end{equation}
But $A \big( K(1) - K(1)^{*} \big) A$ is the sum of $2$ linear
differential operators:
\begin{equation*}
    2 \big( Dm_{0}DA - ADm_{0}D \big) + 2 \big( D^{2}m_{0}A - Am_{0}D^{2}
    \big),
\end{equation*}
which is of order less than $2N+2$ and
\begin{equation*}
    A\big( m_{0}^{\prime}D + Dm_{0}^{\prime} \big)A ,
\end{equation*}
which is of order $4N+1$ unless $m_{0}^{\prime} = 0$ which must be
the case if \eqref{equ:null_criteria} holds. Therefore $m_{0}$ has
to be a constant. Let $\alpha = 2m_{0}\in \mathbb{R}$. Then
\begin{multline*}
    K(m) = \alpha \left\{ 2u_{x}D +
uD^{2} + 2m D^{2}A^{-1} + m_{x}DA^{-1}
    \right\} + \\
    + \beta \left\{ 2u_{x}D^{3} + uD^{4} + 2m D^{4}A^{-1} +
m_{x}D^{3}A^{-1}
    \right\}
\end{multline*}
because $D$ and $A$ commute. The symmetry of the operator $K(m)$
means
\begin{equation}\label{equ:SymmetricOperator}
    \int_{\Ucircle} N\, K(m)M \, dx = \int_{\Ucircle} M\, K(m)N \, dx ,
\end{equation}
for all $m,M,N \in C^{\infty}(\Ucircle)$. Since this last expression
is tri-linear in the variables $m,M,N$, the equality can be checked
for complex periodic functions $m,M,N$. Let $m = Au$, $u=\re^{-\ri
px}$, $M = \re^{-\ri qx}$ and $N = \re^{-\ri rx}$ with $p,q,r \in
\mathbb{Z}$. We have
\begin{multline*}
    \int_{\Ucircle} N\, K(m)M \, dx = \Big[ (2pq^{3} + q^{4})\beta - (2pq +
q^{2})\alpha + \\
    + \Big( (pq^{3} + 2q^{4})\beta - (pq + 2q^{2})\alpha
    \Big)\frac{s_{A}(p)}{s_{A}(q)} \Big] \int_{\Ucircle} \re^{-\ri (p+q+r)x} \, dx \, ,
\end{multline*}
whereas
\begin{multline*}
    \int_{\Ucircle} M\, K(m)N \, dx = \Big[ (2pr^{3} + r^{4})\beta - (2pr +
r^{2})\alpha + \\
    + \Big( (pr^{3} + 2r^{4})\beta - (pr + 2r^{2})\alpha
    \Big)\frac{s_{A}(p)}{s_{A}(r)} \Big] \int_{\Ucircle} \re^{-\ri (p+q+r)x}dx \, .
\end{multline*}

Now we set $p=n$, $q=-2n$, $r=n$ and we must have
\begin{equation}\label{equ:final_equality}
   (24n^{4}\beta -6n^{2}\alpha) s_{A}(n) = ( 6n^{4}\beta -
   6n^{2}\alpha) s_{A}(2n) ,
\end{equation}
if $K(m)$ is symmetric.

If $\beta \neq 0$, the leading term in the left hand-side of
\eqref{equ:final_equality} is $24 \, (-1)^{N} \, a_{2N} \, \beta \,
n^{2N +4}$, whereas the leading term of the right hand-side is $6 \,
(-1)^{N} \, 2^{2N} \, a_{2N} \, \beta \, n^{2N +4}$. Hence, unless
$N=1$, we must have $\beta=0$.

On the other hand, if $\beta=0$, we must have $\alpha
s_{A}(n)=\alpha s_{A}(2n)$, for all $n \in \mathbb{N}^{*}$. Thus
$\alpha=0$ unless $N=0$. This completes the proof.
\end{proof}

\subsection{Hierarchy of first integrals}
\label{subsec:hierarchy}

In view of theorem~\ref{thm:main}, the next step is to find a
hierarchy of first integrals in involution for the vector field
$X_{A}$ where
\begin{equation*}
    A = aI + bD^{2},
\end{equation*}
and $a,b \in \mathbb{R}$ satisfy $a-bn^{2}\ne 0, \forall n \in
\mathbb{Z}$. The vector field
\begin{equation*}
    X_{A}(m) = 2mu_{x} + um_{x} .
\end{equation*}
is bi-Hamiltonian. It can be written as
\begin{equation*}
    X_{A}(m) = P_{m}\, \delta H_{2}(m),
\end{equation*}
where
\begin{equation*}
    H_{2}(m) = \frac{1}{2}\int_{\Ucircle} um \, dx
\end{equation*}
and $P_{m} = mD + Dm$ or as
\begin{equation*}
    X_{A}(m) = Q\, \delta H_{3}(m),
\end{equation*}
where
\begin{equation*}
    H_{3}(m) = \frac{1}{3}\int_{\Ucircle} u(um + q(u))\, dx ,
\end{equation*}
$q(u) = 1/2(au^{2} + bu_{x}^{2})$ and $Q = DA = aD + bD^{3}$.

The problem we get when we try to apply the Lenard scheme to obtain
a hierarchy of conserved integrals is that both Poisson operators
$P_{m}$ and $Q$ are non invertible. However, $Q$ is composed of two
commuting operators, $A$ which is invertible and $D$ which is not.
The image of $D$ is the codimension $1$ subspace,
$C_{0}^{\infty}(\Ucircle)$, of smooth periodic functions with zero
integral. The restriction of $D$ to this subspace is invertible with
inverse $D^{-1}$, the linear operator which associates to a smooth
function with zero integral its unique primitive with zero integral.
Following Lax in \cite{Lax76}, we are able to prove the following
result.

\begin{theorem}\label{thm:Lax}
There exists a sequence $(H_{k})_{k\in\mathbb{N}^{*}}$ of
functionals, whose gradients $G_{k}$ are polynomial expressions of
$u=A^{-1}m$ and its derivatives, which satisfy the \emph{Lenard
recursion scheme}
\begin{equation*}
    P_{m}\, G_{k} = Q\, G_{k+1}.
\end{equation*}
\end{theorem}

\begin{remark}
It is worth to notice, that contrary to the result given by Lax in
\cite{Lax76}, for the KdV equation, the operators $G_{k}$ are
polynomials in $u=A^{-1}m$ and not in $m$. In particular, there are
non-local operators\footnote{Notice that our $m$ corresponds to $u$
in the notations of \cite{Lax76}.}, if $A \ne aI$, for some $a\in
\mathbb{R}$.
\end{remark}

Before giving a sketch of proof of this theorem, let us illustrate
the explicit computation of the first Hamiltonians of the hierarchy.
We start with
\begin{equation*}
    H_{1}(m) = \int_{\Ucircle} m \, dx , \qquad G_{1}(m) = 1.
\end{equation*}
We define $X_{1}$ to be the Hamiltonian vector field of $H_{1}$ for
the Lie-Poisson structure $P_{m}$
\begin{equation*}
    X_{1}(m) = P_{m} \, G_{1}(m) = m_{x}.
\end{equation*}
$X_{1}(m)$ is in the image of $D$ for all $m$ and we can define
\begin{equation*}
    G_{2}(m) = Q^{-1}X_{1}(m) = A^{-1}D^{-1}(m_{x}) = A^{-1}(m)= u
\end{equation*}
which is the gradient of the second Hamiltonian of the hierarchy
\begin{equation*}
    H_{2}(m) = \frac{1}{2}\int_{\Ucircle} mu \, dx .
\end{equation*}
We compute then $X_{2}$, the Hamiltonian vector field of $H_{2}$ for
$P_{m}$
\begin{equation*}
    X_{2}(m) = P_{m} \, G_{2}(m) = 2mu_{x} + m_{x}u = (mu + q(u))_{x} ,
\end{equation*}
where $q(u) = 1/2(au^{2} + bu^{2}_{x})$. $X_{2}(m)$ is in the image
of $D$ for all $m$ and we can define
\begin{equation*}
    G_{3}(m) = Q^{-1}X_{2}(m) = A^{-1}(mu + q(u)) ,
\end{equation*}
which is the gradient of the third Hamiltonian of the hierarchy
\begin{equation*}
    H_{3}(m) = \frac{1}{3}\int_{\Ucircle} u(mu + q(u)) \, dx .
\end{equation*}

So far, we obtain this way a hierarchy of Hamiltonians
$(H_{k})_{k\in\mathbb{N}^{*}}$ satisfying the Lenard recursion
relations for the Euler equation associated to the operator $A$.

\begin{example}[Burgers Hierarchy]
For $A=I$, we obtain explicitly the whole \emph{Burgers hierarchy}
\begin{equation*}
    H_{k+1}(m) = \frac{(2k!)}{2^{k}(k!)^{2}(k+1)} \int_{\Ucircle} m^{k+1} \,
    dx , \qquad (k \in \mathbb{N}).
\end{equation*}
\end{example}

\begin{example}[Camassa-Holm Hierarchy]
For $A=I-D^{2}$, we obtain the \emph{Camassa-Holm hierarchy}. The
first members of the family are
\begin{align*}
    H_{1}(m) & = \int_{\Ucircle} m \, dx = \int_{\Ucircle} u \, dx,\\
    H_{2}(m) & = \frac{1}{2}\int_{\Ucircle} mu \, dx = \frac{1}{2}\int_{\Ucircle} (u^{2} + u_{x}^{2}) \, dx ,\\
    H_{3}(m) & = \frac{1}{2}\int_{\Ucircle} u(u^{2} + u_{x}^{2}) \, dx .
\end{align*}
The next integrals of the hierarchy are much harder to compute
explicitly. One may consider \cite{Len05,Lou05} for further studies
on the subject.
\end{example}

\begin{proof}[Sketch of Proof of Theorem~\ref{thm:Lax}]
The proof is divided into two steps. We refer to \cite{Lax76} for
the details.

\textit{Step 1:} We show by induction that there exists a sequence
of vector fields $G_{k}$, which are polynomial expressions of
$u=A^{-1}m$ and its derivatives and which satisfy
\begin{equation}\label{equ:recurence_relation}
    G_{1} = 1, \qquad PG_{k} = QG_{k+1}, \quad \forall k \in
    \mathbb{N}^{*}.
\end{equation}

\textit{Step 2:} We show that $G_{k}$ is, for all $k$ the gradient
of a function $H_{k}$.

To prove Step 1, we suppose that $G_{1}, \dotsc , G_{n}$ have been
constructed satisfying \eqref{equ:recurence_relation} and we use the
following two lemmas\footnote{The proof of lemma~\ref{lem:Olver} can be found in \cite{Olv93} while the proof of lemma~\ref{lem:Lax} can be found in \cite{Lax76}.} to show that $G_{n+1}$
exists.

\begin{lemma}\label{lem:Olver}
Suppose that $Q$ is a polynomial in derivatives of $u$ up to order
$r$ such that
\begin{equation*}
    \int_{\Ucircle} Q(u) \, dx = 0,
\end{equation*}
for all $u\in \CS$. Then there exists a polynomial $G$ in
derivatives of $u$ up to order $r-1$ such that $Q = DG$.
\end{lemma}

\begin{lemma}\label{lem:Lax}
We have
\begin{equation*}
    \int_{\Ucircle} PG_{n} \, dx = 0
\end{equation*}
for all $n\in\mathbb{N}^{*}$.
\end{lemma}

To prove Step 2, it is enough to show that $G_{k}^{\prime}$ is a
symmetric operator for all $k$, by virtue of
Lemma~\ref{lem:gradient_condition}. We suppose that $G_{1}, \dotsc ,
G_{n}$ are gradients and show first the following result.

\begin{lemma}\label{lem:symmetric_operator}
The operator
\begin{equation*}
    QG_{n+1}^{\prime}(m)Q
\end{equation*}
is symmetric for all $m\in \CS$.
\end{lemma}

We conclude then, like in \cite{Lax76}, that $G_{n+1}^{\prime}(m)$
itself is symmetric. We will give here the details of the proof of
Lemma~\ref{lem:symmetric_operator}, since the proof of the
corresponding result for KdV in \cite{Lax76} is just a direct, hand
waving computation and does not apply in our more general case.

\begin{proof}[Proof of Lemma~\ref{lem:symmetric_operator}]
First, we differentiate the recurrence formula
\eqref{equ:recurence_relation} and we obtain
\begin{equation}\label{equ:recurrence1}
    QG_{n+1}^{\prime}(m) = ad^{*}_{G_{n}} + P_{m}G_{n}^{\prime}(m)
\end{equation}
and
\begin{equation}\label{equ:recurrence2}
    QG_{n}^{\prime}(m) = ad^{*}_{G_{n-1}} +
    P_{m}G_{n-1}^{\prime}(m).
\end{equation}
We multiply \eqref{equ:recurrence1} by $Q$ on the right,
\eqref{equ:recurrence2} by $P$ on the right, and subtract
\eqref{equ:recurrence2} from \eqref{equ:recurrence1}; we get
\begin{equation*}
    QG_{n+1}^{\prime}(m)Q = QG_{n}^{\prime}(m)P_{m} + P_{m}G_{n}^{\prime}(m)Q + ad_{G_{n}}^{*}Q
    - ad_{G_{n-1}}^{*}P_{m} - P_{m}G_{n-1}^{\prime}(m)P_{m}.
\end{equation*}
Using the fact that
\begin{equation*}
    \left(ad^{*}_{u}\right)^{*} = -ad_{u},
\end{equation*}
we get finally
\begin{equation*}
    \left(QG_{n+1}^{\prime}(m)Q\right)^{*} - QG_{n+1}^{\prime}(m)Q =
    Qad_{G_{n}} - P_{m}ad_{G_{n-1}} - ad^{*}_{G_{n}}Q +
    ad^{*}_{G_{n-1}}P_{m}.
\end{equation*}
Using the fact that $Q$ satisfy the following cocycle condition
\begin{equation*}
    Q([u,v]) = ad^{*}_{u}Q(v) - ad^{*}_{v}Q(u)
\end{equation*}
which can be rewritten as
\begin{equation*}
    Qad_{u} = ad^{*}_{u}Q - P_{Q(u)},
\end{equation*}
we get
\begin{equation*}
    \left(QG_{n+1}^{\prime}(m)Q\right)^{*} - QG_{n+1}^{\prime}(m)Q =
    - P_{Q(G_{n})} - P_{m}ad_{G_{n-1}} + ad^{*}_{G_{n-1}}P_{m}.
\end{equation*}
But this last expression is zero because
\begin{equation*}
    P_{m}ad_{v} = ad^{*}_{v}P_{m} - P_{P_{m}(v)}
\end{equation*}
and $Q(G_{n}) = P_{m}G_{n-1}$.
\end{proof}
\end{proof}

\begin{remark}
In the special case where the cocycle $\gamma$ is a coboundary, that
is when the second structure is a \emph{freezing structure}, the
algorithm used to generate a hierarchy of first integrals is known
as the \emph{translation argument principle}~\cite{AK98, KM03}. Let
$H_{\lambda}$ be a function on $\gstar$ which is a Casimir function
of the Poisson structure
\begin{equation*}
    \poisson{\cdot}{\cdot}_{\lambda} = \poisson{\cdot}{\cdot}_{0} + \lambda
    \poisson{\cdot}{\cdot}_{LP}.
\end{equation*}
That is, for every function $F$ one has
\begin{equation*}
    \{H_{\lambda}, F\}_{\lambda} = 0 .
\end{equation*}
Suppose that $H_{\lambda}$ can be expressed as a series
\begin{equation*}
    H_{\lambda} = H_{0}+\lambda H_{1}+\lambda^{2}H_{2}+\cdots
\end{equation*}
Then, one can check that $H_{0}$ is a Casimir function of
$\poisson{\cdot}{\cdot}_{0}$ and that for all $k$, the Hamiltonian
vector field of $H_{k+1}$ with respect to
$\poisson{\cdot}{\cdot}_{0}$ coincides with the Hamiltonian vector
field of $H_{k}$ with respect to $\poisson{\cdot}{\cdot}_{LP}$.
Furthermore, all the Hamiltonians $H_{k}$ are in involution with
respect to both Poisson structures and the corresponding Hamiltonian
vector fields commute with each other. In practice, to obtain such a
Casimir function $H_{\lambda}$, one chooses a Casimir function $H$
of the Poisson structure $\poisson{\cdot}{\cdot}_{LP}$ and then
\emph{translates the argument}
\begin{equation*}
    H_{\lambda} (m) = H(m_{0} + \lambda m) .
\end{equation*}
The above method has been successfully applied to the $\mathrm{KdV}$
equation viewed as a Hamiltonian field on the dual of the Virasoro
algebra.
\end{remark}


\appendix

\section{The Gelfand-Fuks Cohomology}
\label{sec:Gelfan_Fuks}

Gelfand and Fuks \cite{GF68,GR05} have developed a systematic method
to compute the cohomology of the Lie algebra of vector fields on a
smooth manifold. This theory is quite sophisticated. The aim of this
section is to present a computation of the first two cohomological
groups of $\VectS$, using only elementary arguments.

The first difficulty when we deal with infinite dimensional Lie
algebras like $\VectS$ is to define what we call a \emph{cochain},
since a linear or a multilinear map on $\VectS$ may be too vague as
already stated.

\begin{definition}
A $p$-cochain $\gamma$ on $\VectS$ with values in $\mathbb{R}$ is
called \emph{local} if it has the following expression
\begin{equation*}
    \gamma(u_{1}, \dotsc ,u_{p}) = \int_{\Ucircle} P(u_{1}, \dotsc ,u_{p}) \, dx
\end{equation*}
where $P$ is a $p$-linear differential operator.
\end{definition}

It is easy to check that if $\gamma$ is local then $\partial \gamma$
is also local. In the sequel, a cochain on $\VectS$ will always mean
a \emph{local cochain}\footnote{Using a theorem of Peetre
\cite{Pee59}, a local cochain can be characterized by the condition
\begin{equation*}
    \bigcap_{i=1}^{p}Supp(f_{i}) = \emptyset \Rightarrow \gamma
    \left(u_{1}, \dotsc ,u_{p}\right) = 0.
\end{equation*}
}. The associated cohomology is called the \emph{Gelfand-Fuks
cohomology}.

\subsection{The first cohomology group}
\label{subsec:GF1}

A \emph{local} $1$-cochain $\gamma$ on $\VectS$ has the following
expression
\begin{equation*}
    \gamma(u) = \int_{\Ucircle} P(u) \, dx ,
\end{equation*}
where $P$ is a linear differential operator. Integrating by parts,
we can write it as
\begin{equation*}
    \gamma(u) = \int_{\Ucircle} mu \, dx ,
\end{equation*}
where $m\in \CS$ is uniquely defined by $\gamma$.

\begin{proposition}\label{prop:first_cohomology_group}
\begin{equation*}
    H_{GF}^{1}(\VectS;\mathbb{R}) = \set{0}.
\end{equation*}
\end{proposition}

\begin{proof}
If $\gamma$ is a $1$-cocycle, it satisfies the condition
\begin{equation*}
    \gamma([u,v]) = 0 ,
\end{equation*}
for all $u,v$ in $\VectS$. It a very general result that a Lie
algebra which is equal to its commutator algebra has a trivial
$1$-dimensional cohomology group. Indeed, a linear functional which
vanishes on commutators, vanishes everywhere. The proposition is
therefore a corollary of lemma~\ref{lem:commutator_algebra}.
\end{proof}

\subsection{The second cohomology group}
\label{subsec:GF2}

A local $2$-cochain $\gamma$ on $\VectS$ has the following
expression
\begin{equation*}
    \gamma(u,v) = \int_{\Ucircle} P(u,v) \, dx
\end{equation*}
where $P$ is a quadratic differential operator. Integrating by
parts, we can write it as
\begin{equation*}
    \gamma(u,v) = \int_{\Ucircle} uK(v)\, dx ,
\end{equation*}
where $K:\CS \to \CS$ is a linear differential operator
\begin{equation*}
    K = \sum_{k=0}^{n} a_{k}(x)D^{k}
\end{equation*}
which is skew-symmetric relatively to the $L^{2}$-inner product.
This operator is uniquely defined by $\gamma$. If moreover $\gamma$
is a $2$-coboundary, there exists $m\in \g^{*}$ such that $\gamma =
\partial m$, that is
\begin{equation*}
    \gamma(u,v) = - \int_{\Ucircle} m[u,v] \, dx = \int_{\Ucircle} (ad_{u}^{*}\,
    m )v\, dx ,
\end{equation*}
where $u,v \in \g$. We will therefore introduce the following
notation
\begin{equation}\label{equ:coboundary_operator}
    \partial m \, (u) = ad_{u}^{*}\, m = mu_{x} + (mu)_{x} =
    2mu_{x} + m_{x}u,
\end{equation}
to represent the coboundary of the $1$-cochain $m \in \g^{*}$.

\begin{proposition}\label{prop:second_cohomology_group}
The cohomology group $H_{GF}^{2}(\VectS;\mathbb{R})$ is one
dimensional. It is generated by the \emph{Virasoro cocycle}
\begin{equation*}
    \mathrm{vir}(u,v) = \int_{\Ucircle} (u^{\prime}v^{\prime\prime} -
    v^{\prime}u^{\prime\prime}) \, dx .
\end{equation*}
\end{proposition}

\begin{proof}
Let $\gamma$ be a $2$-cocycle and $K$ the corresponding linear
differential operator. The cocycle condition $\partial\gamma = 0$
leads to the following condition on $K$
\begin{equation}\label{equ:cocycle_operator}
    K([u,v]) = ad^{*}_{u}\, K(v) - ad^{*}_{v}\, K(u),
\end{equation}
for all $u,v \in \CS$. Let $w\in\CS$ with zero integral and
$W\in\CS$ a primitive of $w$, we have $w=[1,W]$ and hence
\begin{align*}
    K(w)    & = K([1,W]) \\
            & = ad^{*}_{1}\, K(W) - ad^{*}_{W}\, K(1) \\
            & = K(W)^{\prime} -(2a_{0}W^{\prime} + a_{0}^{\prime}W) \\
            & = \left( a_{1}^{\prime}w + a_{2}^{\prime}w^{\prime} + \dotsb
    +a_{n}^{\prime}w^{(n-1)}\right) + K(w) - 2a_{0}w .
\end{align*}
Therefore we have
\begin{equation*}
    (a_{1}^{\prime}-2a_{0})w + a_{2}^{\prime}w^{\prime} + \dotsb
    +a_{n}^{\prime}w^{(n-1)} = 0
\end{equation*}
for all periodic function $w$ with zero integral which leads to
$2a_{0} = a_{1}^{\prime}$ and $a_{k} = \const$, for $2\leq k \leq
n$. That is, any linear differential linear operator $K$ which
satisfies~\eqref{equ:cocycle_operator} can be written
\begin{equation*}
    K = \partial m + \sum_{k=2}^{n}\lambda_{k}D^{k},
\end{equation*}
where $m$ is a smooth periodic function\footnote{Recall that
$\partial m$ is the linear differential operator defined by
\begin{equation*}
    \partial m \, (u) = ad_{u}^{*}\, m = mu^{\prime} + (mu)^{\prime} =
    2mu^{\prime} + m^{\prime}u.
\end{equation*}
} and the $\lambda_{k}$ are real numbers. Using again
equation~\eqref{equ:cocycle_operator}, we get for all periodic
functions $u,v$
\begin{equation*}
    \sum_{k=2}^{n} \lambda_{k}(uv^{\prime}-vu^{\prime})^{(k)} = 2\sum_{k=2}^{n}
    \lambda_{k}(v^{(k)}u^{\prime}-u^{(k)}v^{\prime}) + \sum_{k=2}^{n}
    \lambda_{k}(v^{(k+1)}u-u^{(k+1)}v) ,
\end{equation*}
which can be rewritten using Leibnitz rule as
\begin{equation*}
    \sum_{k=2}^{n} \lambda_{k} \left\{ \sum_{p=1}^{k-1}C_{k}^{p}(u^{(p)}v^{(k+1-p)}-v^{(p)}u^{(k+1-p)}) + 3(u^{(k)}v^{\prime}-v^{(k)}u^{\prime})
    \right\} = 0.
\end{equation*}
If we fix $v$ and consider this expression as a linear differential
equation in $u$, all the coefficients of that operator must be zero,
and in particular for the coefficient of $u^{\prime}$ we have
\begin{equation*}
    \sum_{k=2}^{n}\lambda_{k}(k-3)v^{(k)} = 0 .
\end{equation*}
Therefore we have $\lambda_{k}=0$ for $k\ne 3$. Since $D^{3}$ is
easily seen to verify \eqref{equ:cocycle_operator}, we can conclude
that every cocycle operator $K$ is of the form
\begin{equation*}
    K = \lambda D^{3} + \partial m
\end{equation*}
for some $\lambda \in \mathbb{R}$ and $m$ in $\CS$. Since every
coboundary operator $\partial m$ is a linear differential operator
of order $1$, $D^{3}$ represent a non-trivial cohomology class,
which ends the proof.
\end{proof}

\bibliographystyle{plain}
\bibliography{biham}
\end{document}